\documentclass[twocolumn]{revtex4}
\usepackage{epsfig}
\usepackage{amssymb}
\usepackage{amsmath}
\usepackage{amsfonts}
\usepackage{graphicx}
\usepackage{mathrsfs}
\usepackage[dvips]{color}
\usepackage{multirow}

\newcommand{\R}{\mathbb{R}}

\newcommand{\fg}{\mathfrak{g}}

\newcommand{\fs}{\mathfrak{s}}

\newcommand{\fz}{\mathfrak{z}}

\newcommand{\bfa}{\mathbf{a}}

\newcommand{\bfr}{\mathbf{r}}

\newcommand{\bA}{\mathbf{A}}

\newcommand{\cP}{\mathcal{P}}

\newcommand{\cT}{\mathcal{T}}

\newcommand{\be}{\begin{equation}}
\newcommand{\ee}{\end{equation}}
\newcommand{\bea}{\begin{eqnarray}}
\newcommand{\eea}{\end{eqnarray}}
\newcommand{\nn}{\nonumber}

\newcommand{\ed}{\end{document}}

\newcommand{\bi}{\begin{itemize}}
\newcommand{\ei}{\end{itemize}}

\newcommand{\bce}{\begin{center}}
\newcommand{\ece}{\end{center}}

\newcommand{\for}{{\mbox{\rm for}}}

\begin{document}

\title{Comment on ``Scattering of light by a parity-time-symmetric dipole\\ beyond the first Born approximation''}

\author{Farhang Loran$^{1}$, Ali~Mostafazadeh$^{2}$, Sema Seymen$^{3}$, and O.~Teoman Turgut$^{3}$}

\address{$^{\!1}$Department of Physics, Isfahan University of Technology,  Isfahan 84156-83111, Iran\\
$^{\!2}$Departments of Mathematics and Physics, Ko\c{c}
University, 34450 Sar{\i}yer, Istanbul, Turkey\\
$^{\!3}$Department of Physics, Bo\u{g}azi\c{c}i University, 34342 Bebek, Istanbul, Turkey
}

\begin{abstract}
In [J.~A.~Rebou\c{c}as and P. A. Brand\~{a}o, Phys.\ Rev.~A {\bf 104}, 063514 (2021)] the authors compute the scattering amplitude for a $\cP\cT$-symmetric double-delta-function potential in three dimensions by invoking the far-zone approximation and summing the resulting Born series. We show that the analysis of this paper suffers from a basic error. Therefore its results are inconclusive. We give an exact closed-form expression for the scattering amplitude of this potential.

\medskip

\end{abstract}

\maketitle

The authors of \cite{RB} consider the scattering problem for a $\cP\cT$-symmetric double-delta-function potential in three-dimensions, 
	\be
	v(\bfr):=\fz_1\delta(\bfr-\bfr_0)+\fz_2\delta(\bfr+\bfr_0),
	\label{double}
	\ee
where $\fz_1=\fz_2^*=-\alpha k^2(\sigma+i\gamma)$ and $\alpha,\sigma$, and $\gamma$ are real parameters, $k$ is the wavenumber, and $\pm\bfr_0$ are the positions of the point scatterers. They substitute the Born series $u(\bfr)=\sum_{n=0}^\infty u_n(\bfr)\alpha^n$ in the Lippmann-Schwinger equation to show that
	\be
	u_n(\bfr)=k^2\int_{\R^3}\chi(\bfr')G(|\bfr-\bfr'|)u_{n-1}(\bfr')d^3 r',~~~n\geq 1,
	\label{eq8}
	\ee
where $\chi(\bfr'):=v(\bfr)/\alpha k^2$, and $G(x):=e^{ikx}/4\pi x$. Then they let $r:=|\bfr|$, $r':=|\bfr'|$, $\hat s:=\bfr/r$, denote the direction of the incident wave vector by $\hat a$, and use $u_0(\bfr)=e^{ik\hat a\cdot\bfr}$ and the far-zone (f.z.) approximation,
	\be
	G(|\bfr-\bfr'|)\sim \frac{e^{ikr}}{4\pi r} e^{-ik\hat s\cdot\bfr'}~~~\for~~~r\gg r',
	\label{eq9}
	\ee
in (\ref{eq8}) to obtain the recursion relation, 
	\begin{align}
	u_n(\bfr)=&\frac{k^2 e^{ikr}}{4\pi r}[(\sigma+i\gamma)
	u_{n-1}(\bfr_0)e^{-ik\hat s\cdot\bfr_0}
	\nn\\
	&+(\sigma-i\gamma)
	u_{n-1}(-\bfr_0)e^{ik\hat s\cdot\bfr_0}],~~~n\geq1.\label{eq11}
	\end{align}
The results of \cite{RB} rely on the authors' solution of this relation. But as we explain below, there is a basic error in their analysis. To determine $u_1(\bfr)$, they substitute (\ref{eq9}) in (\ref{eq8}) and set $n=1$. Because (\ref{eq9}) holds whenever $r\gg r'$, this gives an approximate expression $u_1^{\rm f.z.}(\bfr)$ for $u_1(\bfr)$ which is valid in the f.z., i.e., $u_1(\bfr)\sim u_1^{\rm f.z.}(\bfr)$ for $r\to\infty$. Repeating the same procedure for $n=2$, they express $u_2(\bfr)$ in terms of $u_1(\pm\bfr_0)$ which they calculate by substituting $\pm\bfr_0$ for $\bfr$ in $u_1^{\rm f.z.}(\bfr)$, i.e., set $u_1(\pm\bfr_0)\sim u_1^{\rm f.z.}(\pm\bfr_0)$. This is inadmissible, because $|\bfr_0|$ does not tend to infinity. In general, the iterative solution of (\ref{eq11}) given in the appendix of \cite{RB} is unacceptable, because this equation holds for $r\to\infty$. Therefore, one cannot use it to determine $u_n(\pm\bfr_0)$ even approximately.

There is actually no need for invoking the f.z.\ approximation in treating this problem \cite{demkov,albaverio}. Ref.~\cite{ap-2019} gives the exact solution of the scattering problem for the multi-delta-function potentials, 
	\be
	v(\bfr):=\sum_{n=1}^N\fz_n\delta(\bfr-\bfa_n),
	\label{multi}
	\ee 
in two dimensions, where $\fz_n$ are real or complex coupling constants, and $\bfa_n$ are the positions of the point scatterers. The analysis of \cite{ap-2019} has a straightforward extension to three dimensions. To see this, first we use the notation of Ref.~\cite{RB} to express the Lippmann-Schwinger equation for the potential (\ref{multi}) in the form
	\be
	u(\bfr)=u_0(\bfr)-\sum_{n=1}^N\fz_n G(|\bfr-\bfa_n|)u(\bfa_n).
	\label{LS}
	\ee
Setting $\bfr=\bfa_m$ in this equation, we find a system of linear equations for $u(\bfa_n)$. Because $G(0)=\infty$, the matrix of coefficients of this system has divergent diagonal entries. Therefore, we regularize $G(x)$ and perform a coupling-constant renormalization to remove the singularities. We can do this by a cut-off renormalization or dimensional regularization as outlined in \cite{jackiw} or other renormalization schemes \cite{teo}. In this way, we can set $\bfr=\bfa_m$ in (\ref{LS}) to arrive at
	\be
	\sum_{n=1}^N A_{mn} X_n= e^{ik\bfa_m\cdot \hat a},
	\label{system}
	\ee
where 
	\be 
	A_{mn}:=\left\{\begin{array}{ccc}
	\tilde\fz_n^{-1}+\frac{ik}{4\pi}&\for&m=n,\\[3pt]
	G(|\bfa_m-\bfa_n|)&\for& m\neq n,
	\end{array}\right.
	\label{A=}
	\ee
$\tilde\fz_n$ are the renormalized coupling constants, and $X_n:=\tilde\fz_n u(\bfa_n)$. Solving (\ref{system}) for $X_n$, substituting the result in (\ref{LS}), and noting that the scattering amplitude $\tilde u_s(\bfr)$ is given by 
	\[u(\bfr)\to u_0(\bfr)+\tilde u_s(\bfr)\,\frac{e^{ikr}}{r}~~~\for ~~~r\to\infty,\]
we find 
	\be
	\tilde u_s(\bfr)=-\frac{1}{4\pi} \sum_{m,n=1}^N A_{mn}^{-1}\, e^{ik(\bfa_n\cdot\hat a-\bfa_m\cdot\hat s)},
	\label{scattering-ampl}
	\ee 
where $A_{mn}^{-1}$ are the entries of the inverse of the $N\times N$ matrix $\bA=[A_{mn}]$, and we have also made use of (\ref{eq9}).

For the $\cP\cT$-symmetric double-delta-function potential (\ref{double}), $N=2$, $\tilde\fz_1=\tilde\fz_2^*=-k^2\tilde\alpha(\tilde\sigma+i\tilde\gamma)$, $\tilde\alpha, \tilde\sigma,\tilde\gamma$ are real renormalized parameters, and $\bfa_1=-\bfa_2=\bfr_0$. Subtituting these relations and Eq.~(\ref{A=}) in Eq.~(\ref{scattering-ampl}), we obtain
	\begin{align}
	\tilde u_s(\bfr)=&\frac{1}{2\pi D}\Big[\frac{\fs\cos\xi_--\fg\sin\xi_-}{k^2(\fs^2+\fg^2)}+\nn\\
	&\hspace{1cm}
	\frac{e^{2ikr_0 }\cos\xi_+-2ikr_0\cos\xi_-}{8\pi r_0} \Big],
	\nn
	\end{align}
where 
	\be
	D:=\det\bA=
\frac{2\pi-ik^3\fs}{2\pi k^4(\fs^2+\fg^2)}-
\frac{4k^2r_0^2+e^{4ikr_0}}{64\pi^2 r_0^2},
\nn
	\ee
$\fs:=\tilde\alpha\tilde\sigma$, $\fg:=\tilde\alpha\tilde\gamma$,
$r_0:=|\bfr_0|$, and $\xi_\pm:=k\bfr_0\cdot(\hat a\pm\hat s)$. Notice that the parameters $\fs$ and $\fg$ enter our calculations after we renormalize the bare coupling constants $\alpha\sigma$ and $\alpha\gamma$. Therefore, they may depend on other physical parameters of the problem. 

\vspace{6pt}
\noindent{\em Acknowledgements:}  This work has been supported by the Scientific  and Technological Research Council of Turkey (T\"UB$\dot{\rm I}$TAK) in the framework of the Project No.~120F061 and by the Turkish Academy of Sciences (T\"UBA).
\vspace{12pt}

    

\ed

\bibitem{prl-2009} 
A.~Mostafazadeh,
``Spectral singularities of complex scattering potentials and infinite reflection and transmission coefficients at real energies,''
Phys.\ Rev.\ Lett.~\textbf{102}, 220402 (2009).

\bibitem{pra-2011a} 
A.~Mostafazadeh,
``Optical spectral singularities as threshold resonances,''
Phys.\ Rev.~A {\bf 83}, 045801 (2011).